\documentclass[prl,twocolumn,superscriptaddress]{revtex4-1}
\usepackage{graphicx}
\newcommand{\etal}{{\it et al.}}

\begin{document}

\title{Valence bond phases of herbertsmithite and related copper kagome materials}

\author{M. R. Norman}
\affiliation{Materials Science Division, Argonne National Laboratory, Argonne, IL 60439, USA}
\author{N. J. Laurita}
\affiliation{Department of Physics, California Institute of Technology, Pasadena, CA 91125, USA}
\affiliation{Institute for Quantum Information and Matter, California Institute of Technology, Pasadena, CA 91125, USA}
\author{D. Hsieh}
\affiliation{Department of Physics, California Institute of Technology, Pasadena, CA 91125, USA}
\affiliation{Institute for Quantum Information and Matter, California Institute of Technology, Pasadena, CA 91125, USA}

\begin{abstract}
Recent evidence from magnetic torque, electron spin resonance, and second harmonic generation indicate
that the prototypical quantum spin liquid candidate, herbertsmithite, has a symmetry lower than its x-ray refined
trigonal space group.  Here, we consider known and possible distortions of this mineral class, along with related
copper kagome oxides and fluorides, relate these to possible valence bond patterns, and comment on their
relevance to the physics of these interesting materials.
\end{abstract}

\date{\today}

\maketitle

The nature of the ground state of the near neighbor antiferromagnetic Heisenberg model on a kagome lattice (KAHM) has proven 
to be a challenging problem.
Numerical simulations indicate that a variety of different states have comparable energies, including gapped spin liquids, gapless
spin liquids, and valence bond solids.  This is reflected in the energy spectrum of clusters from exact diagonalization studies,
which shows a dense array of excited states extending down to zero energy \cite{exact}.  In real materials, further richness emerges
due to the presence of anisotropic interactions, such as Dzyaloshinskii-Moriya, as well as longer-range exchange.  In this context,
the lack of observation of an ordered ground state down to 20 mK in herbertsmithite, ZnCu$_3$(OH)$_6$Cl$_2$, a mineral where 
copper ions sit on a perfect kagome lattice, has been a significant result \cite{shores,mendels,norman}.

In reality, though, herbertsmithite is far from perfect.  Single crystals typically contain 15\% of copper ions sitting on interlayer sites
nominally occupied by zinc \cite{han}.  Moreover, despite x-ray refinements of the crystal structure which indicate perfect trigonal symmetry
(R$\bar{3}$m),
magnetic torque and electron spin resonance (ESR) \cite{zorko} find a breaking of the three-fold trigonal axis.  This has been recently
amplified by second harmonic generation (SHG) data, which is consistent with a monoclinic space group that breaks inversion
\cite{laurita}.

To put these results in context, it is first helpful to review known distortions in this mineral class, as well as related materials.
The Cu$_4$(OH)$_6$Cl$_2$ polymorph from which herbertsmithite arises via Zn substitution for Cu, Zn$_x$Cu$_{4-x}$(OH)$_6$Cl$_2$, is 
clinoatacamite with a monoclinic P2$_1$/n space group \cite{grice}.  Upon Zn doping, an intermediate R$\bar{3}$ phase (Zn-paratacamite) is stabilized
between R$\bar{3}$m at high temperatures and P2$_1$/n at low temperatures.  Eventually, the P2$_1$/n phase disappears, and then for
$x$ beyond about 0.34, so does the R$\bar{3}$ phase \cite{welch}.  For Mg-paratacamite, the R$\bar{3}$ phase has been detected up to $x$=0.62 \cite{kampf}.  To understand
the nature of these two structural phases, we employ the crystallographic tools AMPLIMODES \cite{amp} and ISODISTORT \cite{iso}.

The R$\bar{3}$ phase is driven by an F$_2^+$ distortion mode resulting in a quadrupling of the unit cell in the planar directions (Fig.~1, left).
Here, F is equal to (0,$\frac{1}{2}$,1) in hexagonal reciprocal lattice units, and is related to the M ($\frac{1}{2}$,0,0) point of the hexagonal zone
(the difference from M reflects the ABC stacking of layers in the rhombohedral lattice).  The resulting crystallographic distortion
from the F$_2^+$ mode is shown in Fig.~2.  Basically, the interlayer sites (which would nominally be occupied by Zn in stoichiometric
herbertsmithite) divide into two sets, one showing octahedral coordination (1/4 of these sites) and the other a Jahn-Teller distorted
2+2+2 coordination (the remainder).  Around the first type, the atoms on the kagome plane rotate about it.  This is known as a polar
vortex \cite{ramesh} (more formally, an axial toroidal dipole \cite{tugashev}).  The distortion pattern around the other interlayer sites has aspects of this as well, but 
is more complicated.  The actual crystal structure is even more complicated, given the presence of F$_1^+$ and $\Gamma_2^+$ secondary 
modes (Fig.~1, left).  Looking at just
the copper kagome sites \cite{primary}, one finds two crystallographically distinct sites.  This leads to a distribution of Cu-Cu kagome
distances.  The strongest singlet bond (largest Cu-O-Cu bond angle, the superexchange scaling with the bond angle \cite{crawford}) forms a pinwheel pattern, as shown in Fig.~3, left. 
This same pattern is seen in copper kagome fluorides such as Rb$_2$Cu$_3$SnF$_{12}$ \cite{matan}.
Such a pinwheel valence bond pattern has been previously discussed in the KAHM literature given its favorable energetics \cite{pinwheel}.  Note this phase
differs subtly from the so-called diamond valence bond solid, as resonances around diamonds would take one outside of the ground state
manifold \cite{wietek} since those other bonds are not equivalent to the strong bonds.
That is, because of the lattice distortion, there is an
exchange energy cost for a diamond resonance that can 
be estimated as 2(2$J_{12}$-$J_{11}$-$J_{22}$) where $J_{12}$ is the superexchange of the strong bond and 1,2 refer to the two
crystallographically distinct kagome sites.  For a resonance around a pinwheel, the energy cost is 6($J_{12}$-$J^\prime_{12}$) where $J^\prime_{12}$
refers to the 1-2 bond with the smaller Cu-O-Cu bond angle.  Given the linear relation of $J$ with the Cu-O-Cu bond angle for
the bond angle range appropriate to these materials \cite{crawford,rocque,iqbal3},
the estimated cost of a diamond resonance is 0.54$J_{12}$ and a pinwheel resonance 1.45$J_{12}$ for $x$=0.29 \cite{welch}.  These differences
become much smaller as $x$ increases (for Mg-paratacamite at $x$=0.62, they become 0.03$J_{12}$ and 0.02$J_{12}$, respectively \cite{kampf}).

\begin{figure}
\includegraphics[width=0.44\columnwidth]{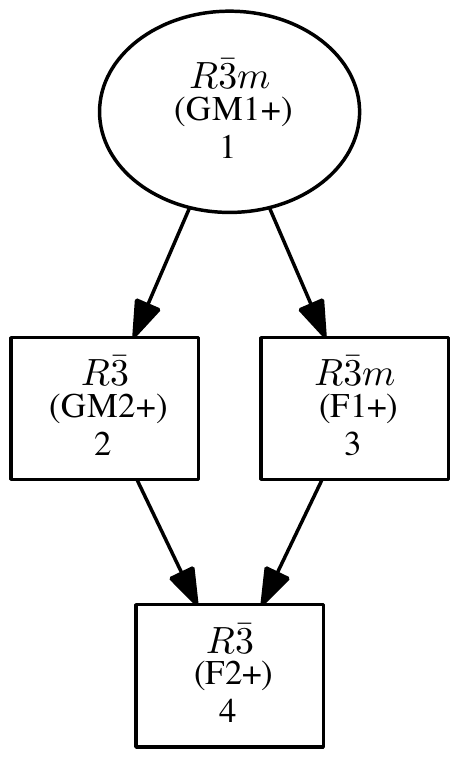}
\includegraphics[width=0.27\columnwidth]{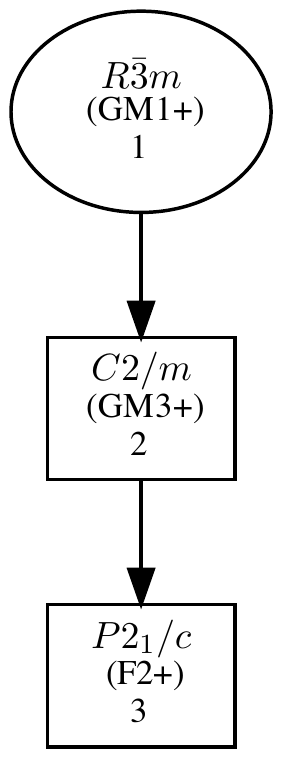}
\caption {Group-subgroup relation leading to left: R${\bar 3}$ and right: P2$_1$/c (equivalent to P2$_1$/n) \cite{bilbao}.
As indicated in these graphs, the primary distortion mode is F$_2^+$, and the secondary modes, arising from
the intermediate groups, are $\Gamma_2^+$ and F$_1^+$ for R${\bar 3}$ and $\Gamma_3^+$ for P2$_1$/c.} 
\label{fig1}
\end{figure}

\begin{figure}
\includegraphics[width=0.9\columnwidth]{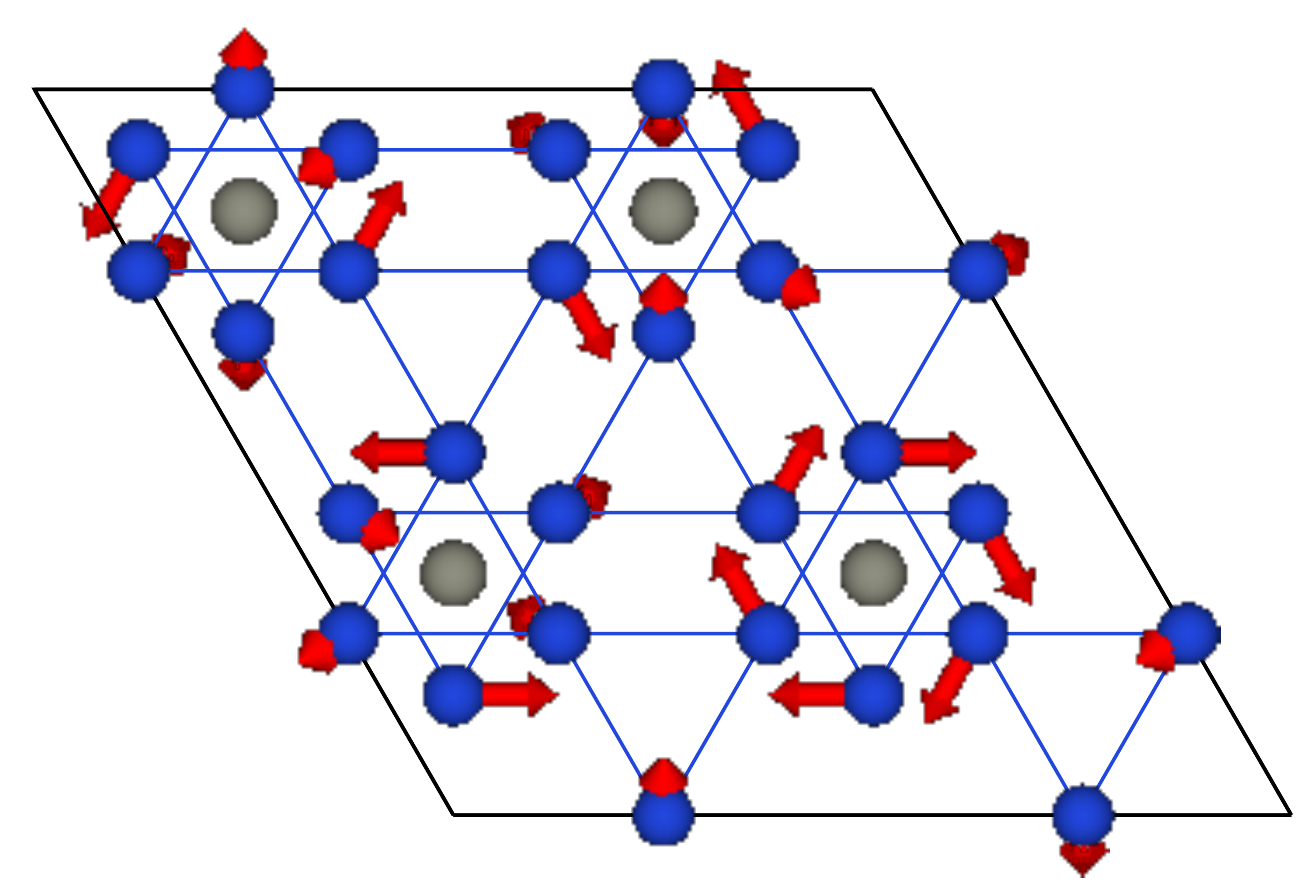}
\caption {F$_2^+$ distortion mode for the R$\bar{3}$ phase of Zn-paratacamite \cite{welch} from AMPLIMODES \cite{amp}
plotted using VESTA \cite{vesta}.  Only the copper/zinc ions are shown in an intersite plane and the two kagome planes
that sandwich it (blue kagome, gray intersite).  Note the vortex-like motion of the kagome coppers
about one of the intersites.  This occurs as well for the oxygen, hydrogen and chlorine ions (not shown).}
\label{fig2}
\end{figure}

The related P2$_1$/n phase seen in clinoatacamite also arises from an F$_2^+$ distortion mode, the difference being due to different
secondary modes ($\Gamma_3^+$ only for P2$_1$/n; Fig.~1, right).  The change in the overall distortion pattern leads to the strong bond now having a herringbone-like
structure (Fig.~3, right).  This has previously been commented on in regards to the Pnma distortion seen in the closely related material, barlowite,
Cu$_4$(OH)$_6$BrF \cite{smaha}.  The distortion in the latter case is an M$_2^+$ mode due to the difference in stacking (AA), the high
symmetry phase being hexagonal (P6$_3$/mmc).  In fact, there is a close analogy between the phase transitions seen in weakly Zn doped
clinoatacamite (R$\bar{3}$m to R$\bar{3}$ to P2$_1$/n) and barlowite (P6$_3$/mmc to P6$_3$/m to Pnma).  In the former case, the primary distortion
mode is F$_2^+$ with secondary modes F$_1^+$ and $\Gamma_2^+$ (R$\bar{3}$) and $\Gamma_3^+$ (P2$_1$/n).  For the latter,
the primary distortion mode is M$_2^+$ with secondary modes M$_1^+$ and $\Gamma_2^+$ (P6$_3$/m) and $\Gamma_5^+$ (Pnma).
These differences again are due to ABC stacking (rhombohedral) versus AA stacking (hexagonal).  This is summarized in Table I.

\begin{figure}
\includegraphics[width=0.45\columnwidth]{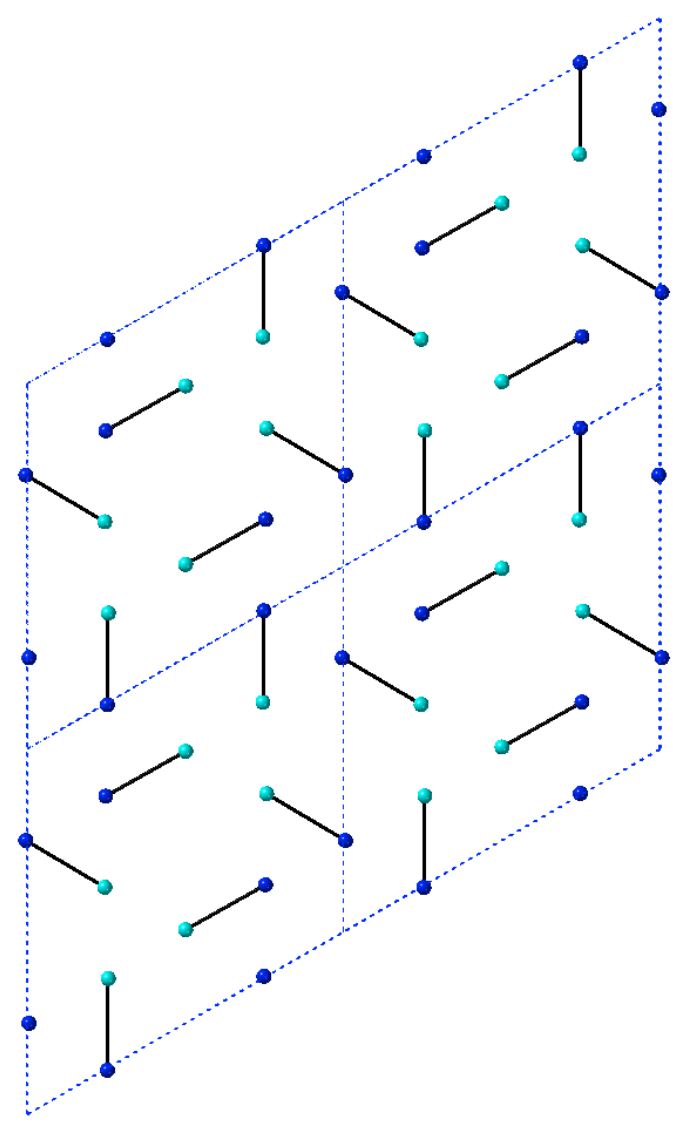}
\includegraphics[width=0.45\columnwidth]{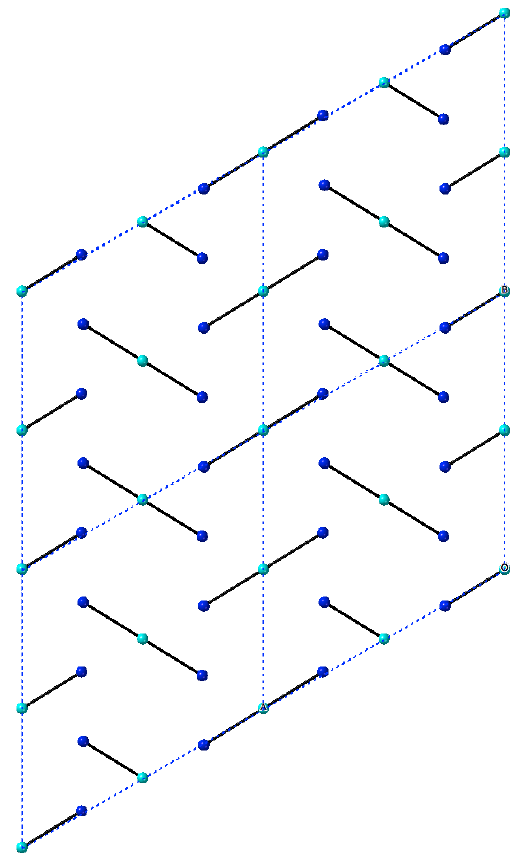}
\caption {Valence bond patterns for left: the R$\bar{3}$ phase (Zn-paratacamite) and right: the P2$_1$/n phase (clinoatacamite).  Similar patterns
are seen for the P6$_3$/m and Pnma phases of barlowite, respectively.  Only the copper ions are shown in a single kagome plane (the two crystallographically distinct
sites are in blue and cyan).}
\label{fig3}
\end{figure}

\begin{table}
\caption{Known valence bond solid (VBS) patterns in copper kagome materials.
low T is the low temperature crystal structure, high T the high temperature one.
mode is the primary distortion mode.  $z$ indicates the net buckling of the copper ions
in the kagome plane along the hexagonal c-axis (in $\AA$).  The high T phase of
Cs$_2$Cu$_3$CeF$_{12}$ is unknown and so was determined from group-subgroup
relations.  Note the large buckling often present in the copper fluorides
as compared to the copper hydroxychlorides.
References are clinoatacamite \cite{malcherek}, 
paratacamite \cite{welch}, barlowite \cite{smaha}, averievite \cite{botana},
Cs$_2$Cu$_3$CeF$_{12}$ \cite{amemiya}, Cs$_2$Cu$_3$ZrF$_{12}$ \cite{reisinger},
Cs$_2$Cu$_3$SnF$_{12}$ \cite{downie} and Rb$_2$Cu$_3$SnF$_{12}$ \cite{downie2}. 
}
\begin{ruledtabular}
\begin{tabular}{llllll}
low T & high T & VBS & material & mode & $z$ \\
\colrule
P2$_1$/n & R$\bar{3}$m & herringbone & clinoatacamite & F$_2^+$ & 0.07 \\
 & & & Cs$_2$Cu$_3$SnF$_{12}$ &	 & 0.07 \\
P2$_1$/c & P$\bar{3}$m1 & herringbone & averievite & M$_2^+$ & 0.00 \\
Pnma & P6$_3$/mmc & herringbone & barlowite (1) & M$_2^+$ & 0.07 \\
R$\bar{3}$ & R$\bar{3}$m & pinwheel & paratacamite & F$_2^+$ & 0.06 \\
 & & & Rb$_2$Cu$_3$SnF$_{12}$ &	 & 0.38 \\
P6$_3$/m & P6$_3$/mmc & pinwheel & barlowite (2) & M$_2^+$ & 0.06 \\
P2$_1$/m & R$\bar{3}$m & zig-zag & Cs$_2$Cu$_3$ZrF$_{12}$  & F$_2^-$ & 0.73 \\
Pnnm & P6$_3$/mmc & stripe & Cs$_2$Cu$_3$CeF$_{12}$  & M$_3^+$ & 3.93 \\
\end{tabular}
\end{ruledtabular}
\end{table}

The detailed temperature dependence of these distortions has been considered by Malcherek \etal~for clinoatacamite \cite{malcherek}
and Welch \etal~for Zn-paratacamite \cite{welch}.  The resulting analysis from AMPLIMODES is shown in Fig.~4.  Despite the expected first-order nature
of the R$\bar{3}$ to P2$_1$/n phase transition, one sees that the F$_2^+$ distortion amplitude goes smoothly through
the transition, and to a good approximation follows a Landau mean-field behavior of $\sqrt{T_{s1}-T}$ where T$_{s1}$ is the upper
transition.  Given the limited data, it is hard to quantify the T dependence of the secondary modes.  Nominally, the amplitude
of the $\Gamma_3^+$ mode should be quadratic in F$_2^+$, but in reality it sets in discontinuously at T$_{s2}$ (lower transition) due to the finite value of F$_2^+$ at T$_{s2}$.

\begin{figure}
\includegraphics[width=0.9\columnwidth]{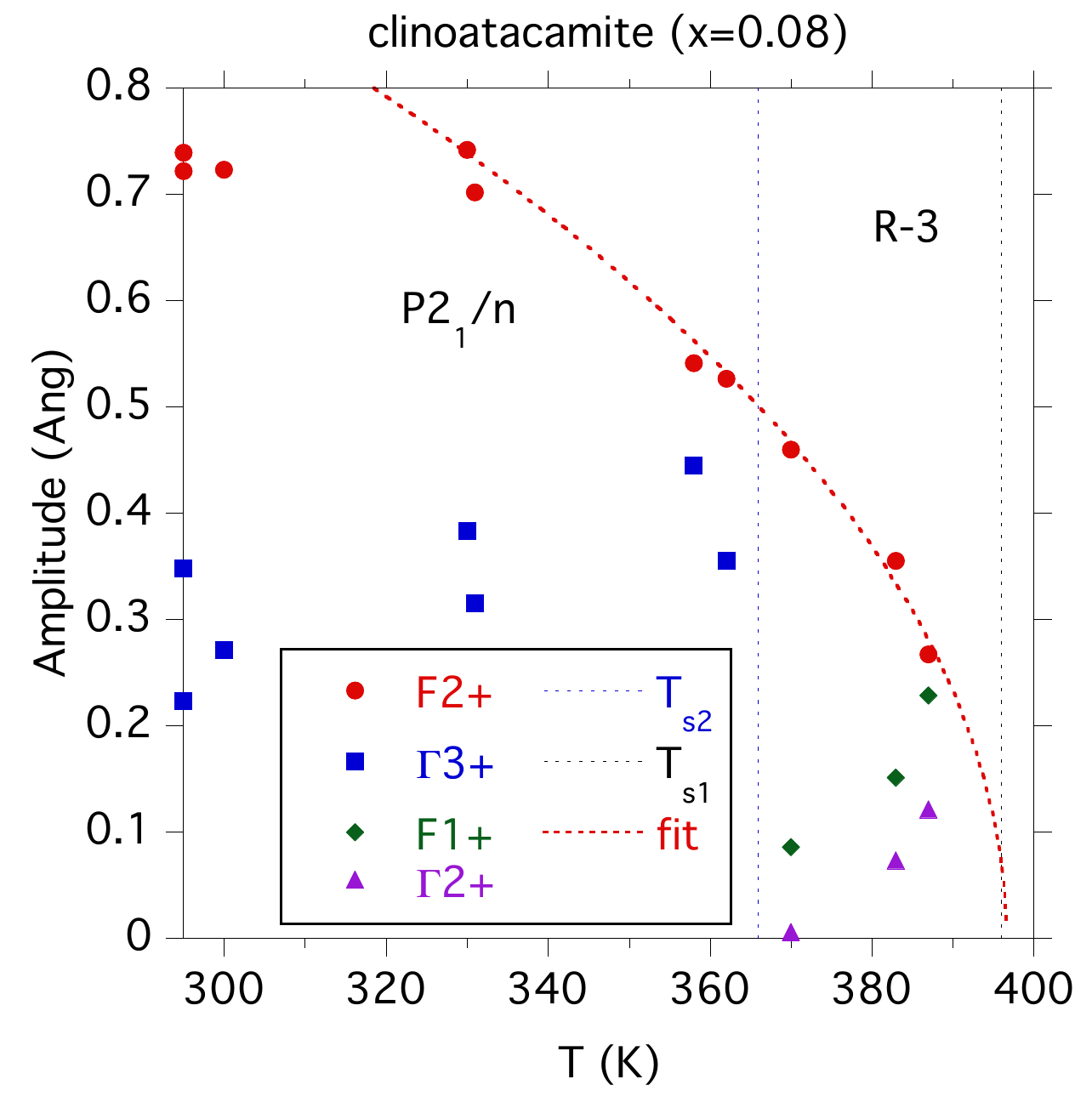}
\caption {Temperature dependence of the distortion mode amplitudes (in~$\AA$) from clinoatacamite \cite{malcherek} generated by AMPLIMODES \cite{amp}.  The red dashed curve is a Landau mean-field fit to the $F_2^+$ mode.}
\label{fig4}
\end{figure}

Returning to the valence bond patterns, a pinwheel pattern is also found in the higher symmetry (P6$_3$/m) version of barlowite \cite{smaha}
(consistent with the above discussed analogy with Zn-paratacamite)
as also listed in Table I.  The known list of patterns can be expanded by considering other materials in the class  A$_2$Cu$_3$BF$_{12}$
where A is an alkali metal and B a 4+ cation \cite{matan2}.  These are also listed in Table I.  In particular, one also finds stripe
phases (Cs$_2$Cu$_3$CeF$_{12}$) and zig-zag phases (Cs$_2$Cu$_3$ZrF$_{12}$).  In all cases in Table I, though, inversion symmetry
is preserved.

This brings us to the SHG data on herbertsmithite \cite{laurita}.  They indicate a point group of either 2 or m.
A likely candidate, then, for the inversion-breaking space group is either Cm or C2.  There are two ways this can happen.
First, by condensing a zone-centered polar mode ($\Gamma_3^-$, Fig.~5).  Possible valence bond patterns are shown in Fig.~6.
Or, by condensing an F-centered
mode (Fig.~7) as shown in Fig.~8 (note, though, that the F$_2^-$ example given in Table I preserves inversion).
Again, these can take the form
of stripes or zig-zags, some of which result from buckled planes.
Note that for illustrative purposes, these patterns are based on the shortest Cu-Cu bonds.  In reality, the strongest singlets will depend on the 
Cu-O-Cu bond angles, meaning oxygen atom displacements need to be considered once they are known.
But one important difference to realize is that for the zone-centered case, one maintains an odd number 
of copper ions (per plane)
in the unit cell.  Therefore, we would anticipate an anisotropic spin liquid in this case rather than a valence bond solid \cite{clark}.
For the zone-boundary modes, though, the unit cell size increases resulting in an even number of copper ions instead, so this would be a valence bond solid.

\begin{figure}
\includegraphics[width=0.45\columnwidth]{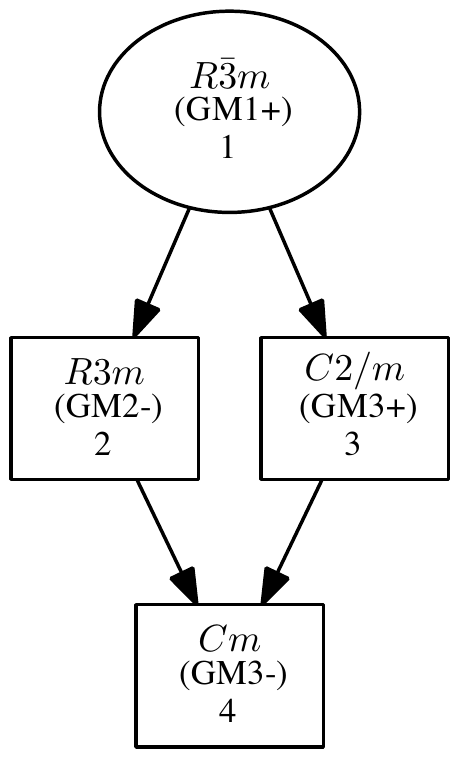}
\includegraphics[width=0.45\columnwidth]{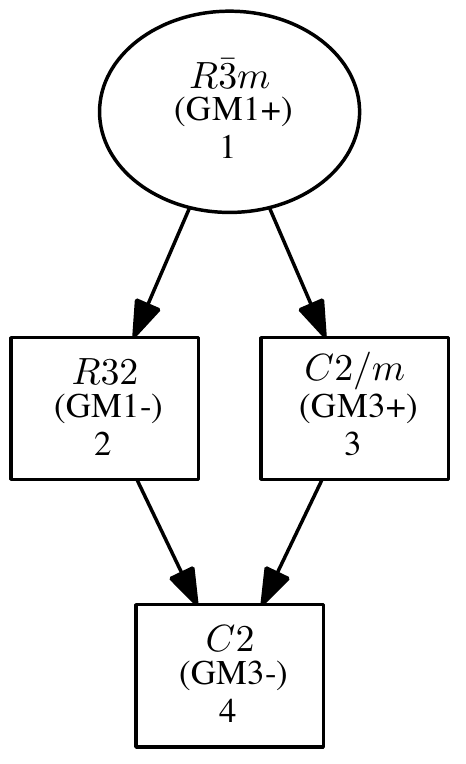}
\caption {Group-subgroup relation leading to left: Cm and right: C2, driven by a $\Gamma$-centered primary mode \cite{bilbao}.
As indicated in these graphs, the primary mode is $\Gamma_3^-$, and the secondary modes, arising from
the intermediate groups, are $\Gamma_2^-$ and $\Gamma_3^+$ for Cm and $\Gamma_1^-$ and $\Gamma_3^+$ for C2.} 
\label{fig5}
\end{figure}

\begin{figure}
\includegraphics[width=0.325\columnwidth]{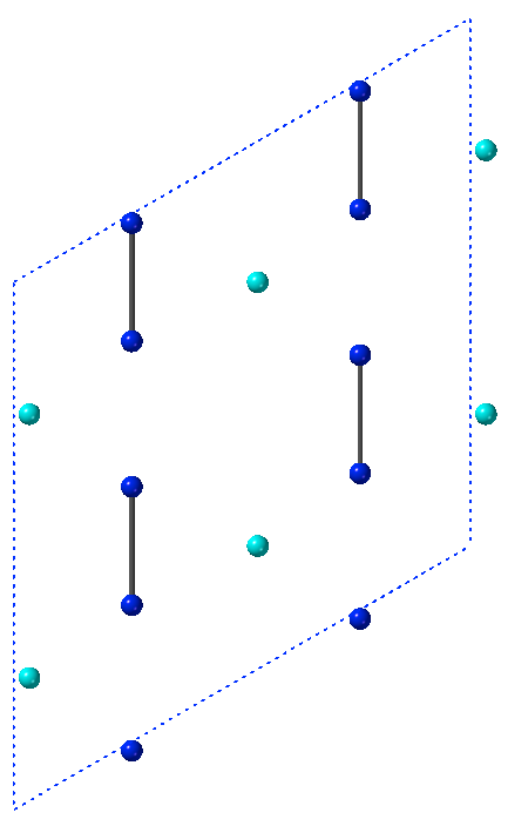}
\includegraphics[width=0.325\columnwidth]{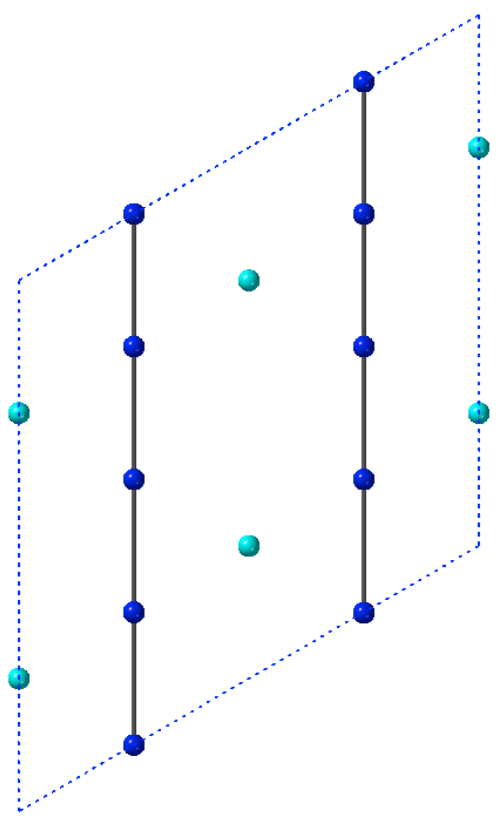}
\includegraphics[width=0.325\columnwidth]{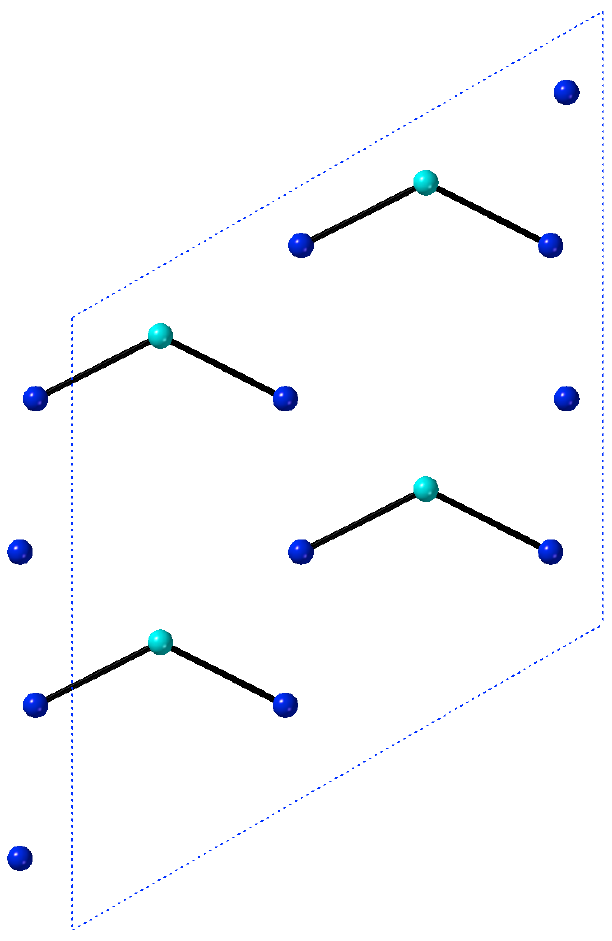}
\caption {Various Cm and C2 VBS patterns from a zone-centered mode ($\Gamma_3^-$) generated by ISODISTORT \cite{iso}.
Left: Cm B$_{u1}$ pattern, Middle: Cm B$_{u2}$ pattern, Right: C2 A$_u$ pattern.  Here, A$_u$ and B$_u$ refer to point group symmetries of the copper kagome
ions.  The Cm A$_u$ pattern (not shown) is similar to the Cm B$_{u1}$ one.  These patterns are based on just copper kagome ion displacements and the shortest Cu-Cu bonds (only the copper ions are shown in a single kagome plane; the two crystallographically distinct
sites are in blue and cyan).
The actual pattern will depend on the Cu-O-Cu bond angles once oxygen ion displacements are known.}
\label{fig6}
\end{figure}

\begin{figure}
\includegraphics[width=\columnwidth]{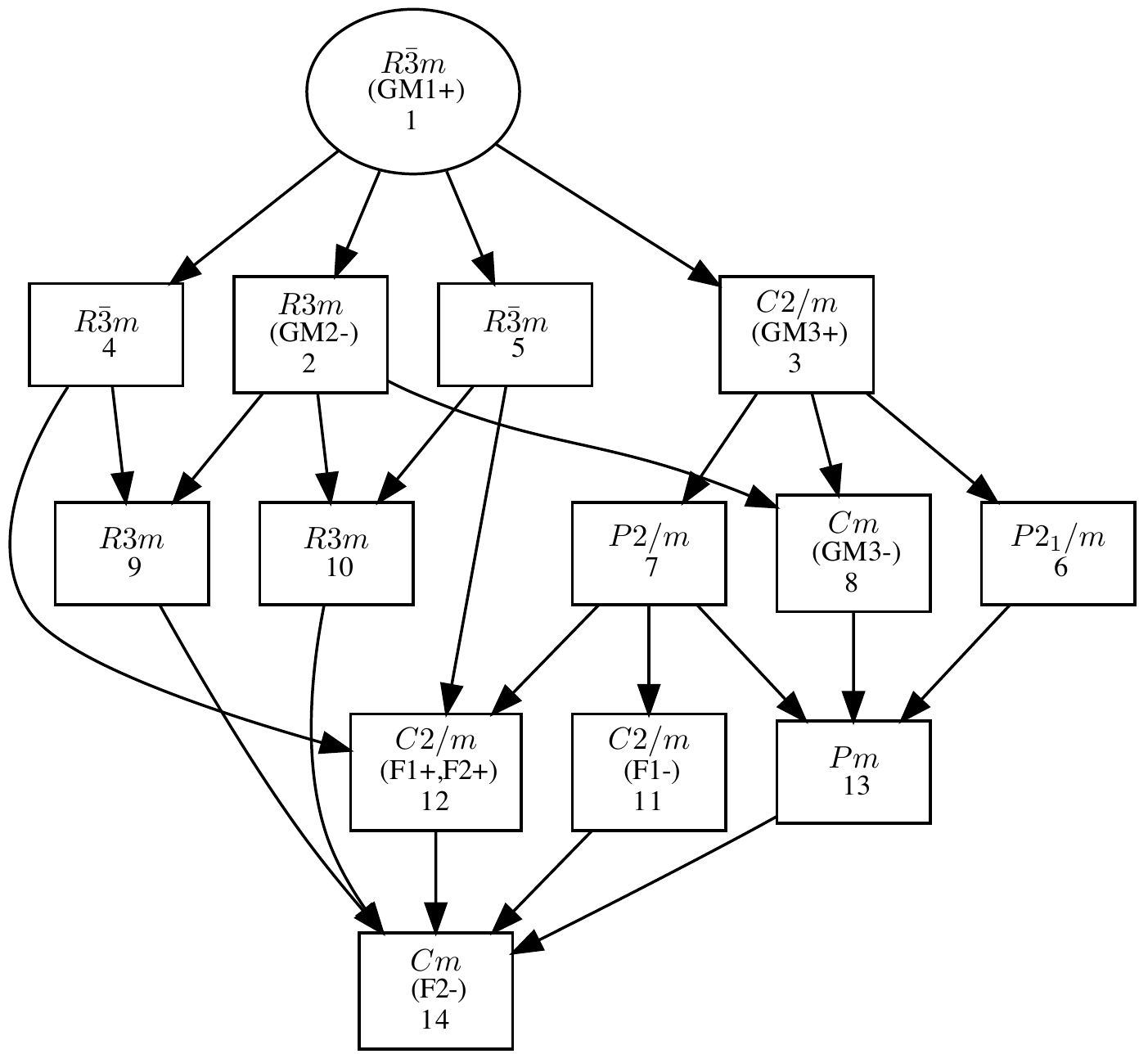}
\includegraphics[width=\columnwidth]{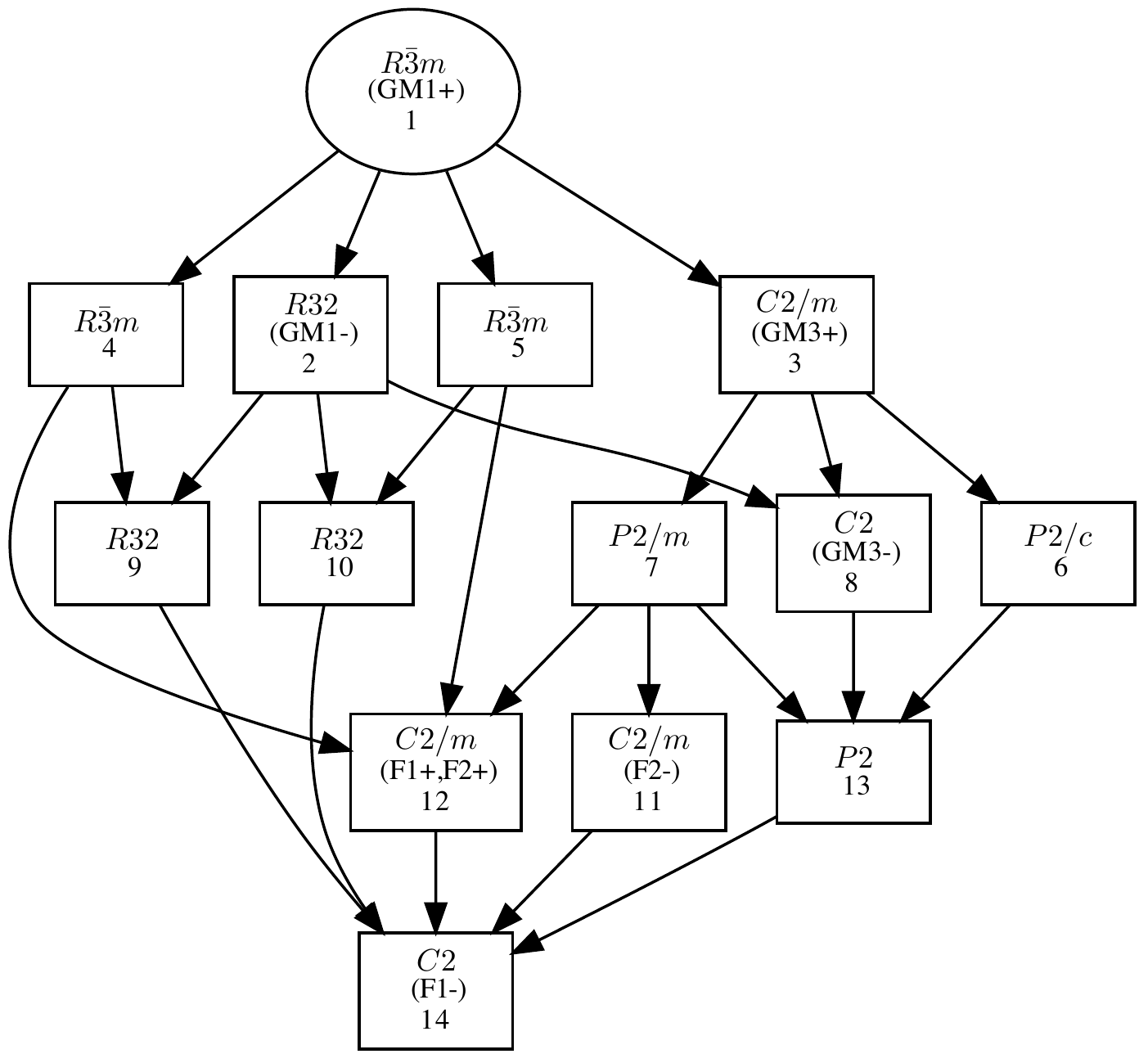}
\caption {Group-subgroup relation leading to top: Cm and bottom: C2, driven by an F-centered primary mode \cite{bilbao}.
As indicated in these graphs, the primary mode is F$_2^-$ for Cm and F$_1^-$ for C2, with a variety secondary modes, arising from
the intermediate groups, that differ for Cm and C2.} 
\label{fig7}
\end{figure}

\begin{figure}
\includegraphics[width=0.45\columnwidth]{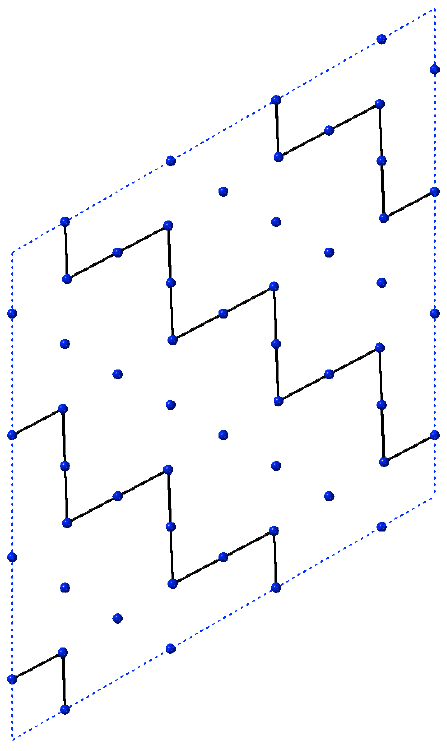}
\includegraphics[width=0.45\columnwidth]{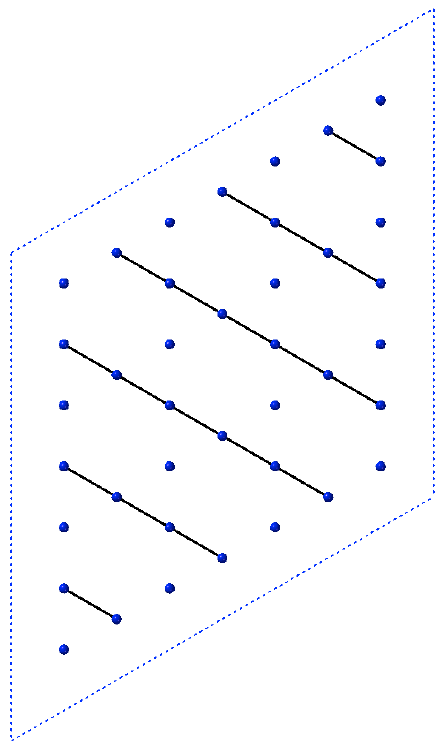}
\caption {Two of the Cm VBS patterns from a zone-boundary mode (F$_2^-$) generated by ISODISTORT \cite{iso}.
Left: one of the two B$_{u1}$ patterns, Right: one of the two B$_{u2}$ patterns.  The other two patterns are dimer patterns similar to Fig.~6a.
These patterns are based on just copper kagome ion displacements and the shortest Cu-Cu bonds (only the copper ions are shown in a single kagome plane;
there are seven crystallographically distinct sites).
The actual pattern will depend on the Cu-O-Cu bond angles once oxygen ion displacements are known.}
\label{fig8}
\end{figure}

One interesting point about the Cm and C2 space groups is that they are in general ferroelectric, with the polar axis along the 2-fold (hexagonal
b axis) for C2, and perpendicular to this axis for Cm.  Ferroelectric behavior has been claimed for the cobalt analog of clinoatacamite, with
a proposed structural distortion of R3m \cite{cobalt}.
Indications of ferroelectric-like behavior has also been seen in averievite \cite{dressel}, where a transition from
the intermediate P2$_1$/c phase to a lower temperature phase of unknown symmetry has been observed \cite{botana}.
This brings up the question of spin-lattice coupling.  The SHG signal in herbertsmithite follows the predicted temperature dependence of
the spin-spin correlator for a kagome lattice \cite{laurita}.  This is consistent with the temperature dependence of phonon linewidths \cite{sushkov}.  Moreoever, the phonon frequencies shift \cite{sushkov},
also indicative of spin-lattice coupling as has been studied extensively in
pyrochlores \cite{pyro,spinel}.  
The idea here is that the superexchange $J$ is sensitive to distortions given its dependence on
the Cu-O-Cu bond angle, harking back to early work by Baltensperger \cite{balten}, with the distortion occurring if the gain in exchange
energy from increasing the bond angle exceeds the elastic cost of the lattice distortion.
Spin-lattice couplings have been quantified in clinoatacamite
using Raman data \cite{liu}.  Ultimately, they can lead to multiferroic behavior, as observed in the distorted kagome material KCu$_3$As$_2$O$_{7}$(OH)$_3$
\cite{nilsen}.

Finally, what does all of this have to do with the KAHM?  Density matrix renormalization group simulations have indicated that the ground state is a melted version of
a twelve-site diamond valence bond solid, closely related to the pinwheel pattern \cite{DMRG}.  This has been further investigated by more recent
numerical work \cite{rouso},  though related numerical simulations favor a Dirac spin liquid instead \cite{ran,iqbal2,DSL}.  Small perturbations could certainly stabilize a valence 
bond solid \cite{huh,iqbal,hwang}, or an anisotropic spin liquid \cite{clark,repellin}.  Given the above results, such models should be further explored to understand the
rich physics of the Heisenberg model on a kagome lattice and its material realizations.

M.R.N. was supported by the Materials Sciences and Engineering Division, Basic Energy Sciences, Office of Science, U.S. Dept.~of Energy.
N.J.L. acknowledges partial support from an Institute for Quantum Information and Matter Postdoctoral Fellowship. 
D.H. acknowledges support from an ARO PECASE award W911NF-17-1-0204.


\begin{thebibliography}{99}
\bibitem{primary}
For simplicity, in this paper we focus on copper ion displacements.  Oxygen ion displacements play an equally important role as indicated
later.  Which one dominates does not affect the definition of primary versus secondary modes, which is set by the group-subgroup relations
and resulting Landau free energy, though the nature of the ion displacements depends critically on the symmetry of these modes.
\end{thebibliography}
\end{document}